\documentclass[11pt]{llncs}
\usepackage{epsfig}
\usepackage{amssymb}
\usepackage{amsmath}

\begin{document}

\title{On the undecidability of the Panopticon detection problem}

\author{V. Liagkou \inst{1,4}  
\and P.E. Nastou \inst{5} \and P. Spirakis \inst{2} \and Y.C. Stamatiou\inst{1,3}}

\institute{Computer Technology Institute and Press - ``Diophantus'', University of Patras Campus, 26504, Greece \and
Department of Computer Science, University of Liverpool,
UK and Computer Engineering and Informatics Department, University of Patras, 26504, Greece \and
Department of Business Administration, University of Patras, 26504, Greece \and
University of Ioannina, Department of Informatics and Telecommunications, 47100 Koatakioi Arta, Greece  \and
Department of Mathematics, University of the Aegean, Applied Mathematics and Mathematical Modeling Laboratory, Samos, Greece
\\ e-mails: liagkou@cti.gr, pnastou@aegean.gr, P.Spirakis@liverpool.ac.uk, stamatiu@ceid.upatras.gr}

\maketitle


\begin{abstract}
The Panopticon (which means ``watcher of everything'') is
a well-known structure of continuous surveillance and discipline proposed by
Bentham in 1785. This device was, later, used by Foucault
and other philosophers as a paradigm and metaphor for the study
of constitutional power and knowledge as well as a model of individuals' deprivation of freedom.
%
%
Nowadays, technological achievements have given rise
to new, non-physical (unlike prisons), means of constant surveillance 
that transcend physical boundaries.
%
%
This, combined with the confession of some governmental institutions that they actually
collaborate with these Internet giants to collect or deduce
information about people, creates a worrisome situation of several co-existing
Panopticons that can act separately or in close collaboration.
Thus, they can only be detected and identified
through the expense of (perhaps considerable) effort.
In this paper we provide a theoretical framework for
studying the detectability status of Panopticons that fall under two theoretical, but not unrealistic, definitions.
We show, using Oracle Turing Machines, that detecting modern day, ICT-based, Panopticons
is an undecidable problem. Furthermore, we show
that for each sufficiently expressive formal system, we can effectively
construct a Turing Machine for which it is impossible to prove, within the formal system,
either that it is a Panopticon or it is not a Panopticon.
\keywords{Formal Methods \and Security \and Privacy \and Undecidability \and
Panopticon \and Turing Machine \and Oracle Computations}
\end{abstract}

\section{Introduction}

In 1785, the English philosopher and social theorist
Jeremy Bentham (see~\cite{Bentham})
envisaged an, admittedly, unprecedented (for that era) institutional
punishment establishment, the {\em Panopticon}. The architecture of this establishment
consisted of a circular building dominated by an ``observation tower''
in the center of which a
single guard was continuously watching the inmates,
imprisoned in cells arranged around the circular building.
Standing on the observation tower, the prison’s inspector
was able to observe the interior of the cells at any time. Moreover,
the prisoners, themselves, could never be able to see
the inspector, who remained for ever ``invisible'' to them.

In the \textquotesingle70s, Foucault studied, deeply, Bentham’s concepts, pointing to the Panopticon as a generic
model that denotes a way of defining and discussing {\em power}
relations in terms of their impact on people’s everyday life. He, also, described the Panopticon as a
mechanism of power enforcement that is reduced to its ideal form:
``a figure of political technology that may and must be detached from any specific use'' (\cite{Foucault77}).
According to Foucault, disciplinary power was increasingly
permeating, in his era, the social body in schools, factories, hospitals,
asylums and military barracks constituting ``new prison regimes'' in the emergent capitalist society.
His own conceptualization and usage of Panopticon has allowed for several more or less metaphorical, yet extensive,
usages of the {\em all-seeing} abilities
that the ``panopticized'' state of affairs offers in diverse core field areas. This was, in retrospect,
an indirect reference to the modern day, information technology based, surveillance and people monitoring
operations conducted, openly or covertly, by several agencies and organizations worldwide.

%
Nowadays, our ``digital selves'' and personal
digital assets and information
transcend physical bounds and, literally, are diffused over the vast, uncontrollable, Internet territory.
This dispersion of our digital assets and personal information provides unlimited
opportunities for massive data collection and surveillance of our daily actions by
state agencies, intelligence institutions and Internet service providers. Several
voices exist that equate this situation to an information era
{\em Panopticon} or {state of massive surveillance} concept.
In particular, contrary to the ``brick-and-mortar''
Panopticon of Bentham, the surveillance actors often remain invisible and covert since modern
surveillance methods and devices are hard to detect, unlike the classical Panopticon physical structure
of whose existence and objectives all its subjects are aware.

In this paper we investigate the complexity of detecting Panopticons using the
{\em Turing Machine} formalism of an effective computational procedure.
We provide two different, but not unrealistic, theoretical models of a Panopticon
and show that there is no algorithm, i.e. Turing machine, that can detect, systematically, all Panopticons under these two definitions.
In other words, detecting Panopticons, at least the ones that fall under
these two plausible definitions, is an undecidable problem, in principle.
%

More specifically, the first formal model we examine studies Panopticons whose Panopticon behaviour
is manifested through the {\em execution} of states (actions) that belong in a specific set of states that characterizes
Panopticon behaviour. In some sense, since the focal point of this model is the {\em execution} of states
of a particular type, the model captures the {\em visible behaviour} of the Panopticon, according to the action
ir perform, and, thus we call this model {\em behavioural}.

The second formal model focuses on the {\em impact} or {\em consequences} of the actions of the Panopticon
and not the actions themselves. In particular, this model captures an essential characteristic of Panopticons,
that of acquiring, rather, {\em effortlessly} information through {\em surveillance} and {\em eavesdropping}.
We model this characteristic using {\em Oracle} Turing Machines with the oracle having
the role of information acquired ``for free'' based on surveillance
and/or eavesdropping actions. This model is in some sense based on the information that a Panopticon deduces
using ``free'' information and, thus, we call it {\em deductive}.
Essentially, this model focuses on the {\em semantics} of a Turing Machine, i.e. outcomes of
operation, while the first model
focuses on the {\em syntax}, i.e. definition, of a Turing Machine.

Last, we show that for any formal system, we can demonstrate a Turing Machine
whose Panopoticon status, under the second formal model, cannot be proved within the formal system.
That is, no proof can be produced by the formal system
that this Turing Machine is a {\em Panopticon} and no proof
that it is not a {\em Panopticon}. In other words, given any formal system, one can provide a
procedure that generates a Turing Machine which is impossible to
have its Panopticon or non-Panopticon status within the formal system.

\section{Definitions and notation}
In this section we briefly state the relevant definitions and notation that will be used in the subsequent sections. We,
first, define a simple extension of a Turing Machine, following the notation in~\cite{HopUll79}.

\noindent
\begin{definition} [Turing machines] A Turing machine is an octuple, defined as
$M\mathrm{=}(Q, Q_{\mbox{\small{pan}}}, \mathit{\Sigma},\mathit{\Gamma},\delta ,q_0,B,F)$ where $Q$ is a finite set of
normal operation states, $\mathit{\Gamma}$ is a finite set called the tape alphabet, where $\mathit{\Gamma}$ contains a special symbol $B$ that represents a blank, $\mathit{\Sigma}$ is a subset of $\mathit{\Gamma}\mathrm{-}\left\{B\right\}$ called the input alphabet, $\delta $ is a partial function from $Q\mathrm{\times }\mathit{\Gamma}$ to $Q\mathrm{\times }\mathit{\Gamma}\mathrm{\times }\left\{L,R\right\}$ called the transition function, $q_0\mathrm{\in }Q$ is a distinguished state called the start state, $F \subset Q$ is a set of final states, and $Q_{\mbox{\small{pan}}} \subset Q$, $Q_{\mbox{\small{pan}}} \cap F = \emptyset$, is a distinquished set of states linked to Panopticon behaviour. We assume that transitions from states in $Q_{\mbox{\small{pan}}}$
do not change the Turing machine's tape contents, i.e. they
are purely interactions with the external environment of the Turing machine and can affect only the environment.
\label{def-TM}
\end{definition}
Notation-wise, given $M$ we denote by $<M>$ its {\em code},
i.e. an encoding of its description elements as stated in Definition~\ref{def-TM}
using any fixed alphabet, usually the alphabet $\{0,1\}$ (binary system).
The details can be found in, e.g.,~\cite{Evans,HopUll79} but they
are inessential for our arguments.


One of the main outcomes of Turing's pioneering work~\cite{Turing36} was
that there exist problems that Turing machines cannot solve. The first, such, problem was
 the, so called, {\em Halting problem} (see, also,~\cite{Davis00} for an excellent historic account):

\medskip
\noindent {\bf The Halting Problem}

\noindent {\bf Input:} A string $x = <M,w>$ which is actually the encoding (description) of a Turing machine $\mathrm{<}$M$\mathrm{>}$ and its input $w$.

\noindent {\bf Output:} If the input Turing $M$ machine halts on $w$, output True. Otherwise, output False.
\medskip

The language corresponding to the Halting problem is $L_u = \{<M,w> | w \in L(M)\}$. In other words, the language $L_u$ contains all possible {\em Turing machine-input} pair encodings $<M,w>$ such that $w$ is accepted by $M$. This is why $L_u$ is also called {\em universal language} since the problem of deciding whether a given Turing machine $M$ accepts a given input $w$ is equivalent to deciding whether $<M,w> \in L_u$. The  language $L_u$ was the first language proved to be non-recursive or undecidable by Turing.

In order to discuss Panopticons, we need an important variant of Turing machines,
called {\em oracle} Turing Machines. Such a machine
has a special tape on which it can write queries to which they
obtain the answer instantaneously in one step, no matter what query it is.
%
%

This type of Turing Machines was, first, discussed, briefly, by Turing himself in~\cite{Turing39} under the
name {\em o-machine}. Post developed further this concept in a series of papers~\cite{Post43,Post44,Post48}
and his collaboration with Kleene in~\cite{Kleene-Post54} resulted
to the definition that is used today in computability theory.

Below, we give a formal definition of an Oracle Turing Machine:

\begin{definition} [Oracle Turing Machine]
Let $A$ be a language, $A\mathrm{\subseteq }{\mathit{\Sigma}}^{\mathrm{*}}$. A Turing machine with oracle $A$ is a single-tape Turing machine with three special states $q_{\mathrm{?}},q_y$ and $q_n$. The special state $q_{\mathrm{?}}$ is used to ask whether a string is in the set$\mathrm{\ }A$. When the Turing machine enters state $q_{\mathrm{?}}$ it requests an answer to the question: ``Is the string of non-blank symbols to the right of the tape head in $A$?{'}{'} The answer is provided by having the state of the Turing machine change on the next move to one of the states $q_y$ or $q_n$. The computation proceeds normally until the next time $q_{\mathrm{?}}$ is reached, at which point the Turing machine requests another answer from the oracle.
\end{definition}
With respect to notation, we denote by $M^A$ the Turing machine $M$ with oracle $A$.
Also, a set (language) $L$ is recursive with respect to $A$ if $L=L(M^A)$ for some Turing machine $M^A$ that {\em always} halts
while two oracle sets (languages) are called {\em equivalent} if each of them is recursive in the other (see~\cite{HopUll79}).


\section{Our contributions}
\label{contrib}

Our approach is different for each of the two Panopticon models we propose since they
are of a different nature, i.e syntactic (for the behavioural model) vs. semantic (for the deductive model).

For the behavioural model, we provide a simple adaptation of Cohen's piooneering formal model of a {\em virus} and
prove a Panopticon detection impossibility result much like Cohen's result for virus detection.

For the deductive model, we follow a completely different approach
using Oracle Turing Machines and a technique that
can be applied to prove undecidabililty results for this type of machines.
More specifically, in Chapter 8 of~\cite{HopUll79} a technique from~\cite{HopUll68}
is presented that establishes an hierarchy of undecidable problems for Oracle Turing Machines.
In particular, The technique targets the
oracle set $S_{\mathrm{1}}\mathrm{=}\left\{\mathrm{<}M\mathrm{>|}L(M)\mathrm{=}\mathrm{\emptyset }\right\}$,
with $\mathrm{<}M\mathrm{>}$ denoting the encoding of Turing machine $M$, as we discussed before.
Then, the sets
$S_{i\mathrm{+1}}\mathrm{=}\left\{\mathrm{<}M\mathrm{>|}L^{S_i}(M)\mathrm{=}\mathrm{\emptyset }\right\}$ can be, recursively, defined and the following can be proved (see~\cite{HopUll68,HopUll79}):
\begin{theorem}
The membership problem for TM's without oracles is equivalent to $S_1$ (i.e. $L_u$ is equivalent to $S_1$).
\label{T1}
\end{theorem}

\begin{theorem}
The problem of deciding whether $L(M) = \Sigma^*$ is equivalent to $S_2$.
\label{T2}
\end{theorem}
Our first contribution is to propose a plausible Panopticon model which incorporates the
{\em information deduction} element of its behaviour (see Definition~\ref{pan-oracle}). 
We accomplish this as follows: information deduction takes place
whenever the Turing machine under scrutiny for Panopticon
behaviour produces a {\em completely new}
information set given a set of fixed, finitely many,
already {\em known} information sets. This set models the information
that the Panopticon {\em already knows}
through surveillance and observation, without (usually) expending considerable
effort since it, merely, intercepts or eavesdrops information.

More formally, let $N_i=\left\{L^i_1,L^i_2,\dots ,L^i_k\right\}$ be a set of
recursively enumerable languages, for some fixed
integer $k \geq 1$, such that $\emptyset \notin N_i$ for all $i$. Also, let $M^i_1, M^i_2, \dots, M^i_k$ Turing machines that, correspondingly, accept these languages.
These Turing machines and their corresponding languages model the fixed,
finitely many, information sets already known to the Panopticon.
%
%
%
We, also, say that a set is \textit{disjoint} from a collection of sets if it is disjoint from all the sets in the collection. set is \textit{disjoint} from a collection of sets if it is disjoint from all the sets in the collection. 
%
%

We will, now, define the oracle set
$S_1=\left\{<M>|L(M) \mathrm{\ is\ disjoint\ from}\ N_1\right\}$,
with $<M>$
denoting the encoding of Turing machine $M$,
and, recursively, 
in analogy with~\cite{HopUll68,HopUll79},
the sets $S_{i+1} = \left\{<M>|L(M^{S_i}) \mathrm{\ is\ disjoint\ from}\ N_{i+1}\right\}$.
The sets
$S_1$ and $S_2=\left\{<M>|L(M^{S_1})\ \mathrm{is\ disjoint\ from}\ N_2\right\}$,
in particular, are central to our approach.

Based on this framework, in Section~\ref{ded} we
prove two theorems analogous to Theorems~\ref{T1} and~\ref{T2}
on the undecidability of the problem of detecting a deductive Panopticon.
The first one, Theorem~\ref{1st}, is focused on the weaker form of the deductive Panopticons,
related to the set $S_1$, while the more powerful one, based
on oracle computation for ``free'' information gathering, related to the set $S_2$,
is handled by Theorem~\ref{2nd}.
In particular,
in Theorem~\ref{1st} we prove that $L_u$ is equivalent
to $S_2$ and in Theorem~\ref{2nd} we prove that the problem of
whether $L(M) = \Sigma^*$ is equivalent to $S_2$.

Finally, in Section~\ref{G}
we show that for any sufficiently expressive formal system ${\cal F}$,
such as Set Theory, we can effectively construct a Turing Machine which is impossible
to classify it as a Panopticon or non-Panopticon within ${\cal F}$. In other
words no formal system is powerful enough
so that given any Turing Machine, it can provide either a proof that it is a Panopticon
or a proof that it is not a Panopticon.

Before continuing, we should remark
that the essential element of the proposed definition of deductive Panopticons
is that the oracle consultations model the ``effortless'', through surveillance, interception or eavesdropping,
information gathering by Internet surveillance agencies and organizations. In this
context, the sets $S_{i+1}$ define an infinite {\em hierarchy} of deductive Panopticons
in which a Panopticon whose accepted language belongs in $S_{i+1}$ operates
by consulting a (weaker) lower-level Panopticon whose language belongs in $S_{i}$, with
the weakest Panopticons being the ones whose accepted languages belong in $S_1$. These Panopticons
do not have oracle consultations or effortless information gathering capabilities.

\section{The Panopticon detection problem}

\label{theorems}

Formal proofs about the impossibility of detecting, in a {\em systematic}
(i.e. algorithmic) and {\em general} way, malicious entities, such as the Panopticons
in our case, already exist for a long time for a very important
category of such entities, the {\em computer viruses} or {\em malware} in general. 

In Cohen's pioneering work (see~\cite{Cohen85,Cohen87})
a natural, {\em formal}, definition of a virus is provided based on Turing machines.
Specifically, Cohen defined a virus to be a program, or Turing machine,
that simply copies itself to other programs,
or more formally, injects its transition function into other Turing machines'
transition functions (see~\ref{def-TM}) replicating, thus, itself indefinitely.
Then, he proves that $L_u$ reduces to the problem of deciding whether
a given Turing Machine behaves in this way proving
that detecting viruses is an undecidable problem.

Following Cohen's paradigm, we will propose two, rather restricted (so as to be amenable to a theoretical analysis) but reasonable and precise definitions of a Panopticon. We, first, define the {\em behavioural Panopticons}:

\begin{definition} (Behavioral Panopticons)
A Panopticon is a Turing machine that when executed will demonstrate a specific, recognizable,
behaviour particular to Panopticons manifested by the {\em execution} (not simply the
existence in the Turing machine's description) of a sequence of actions, e.g. it
will publish secret information about an entity, it will download information illegally etc., actions reflected by reaching, during its operation, states in the set $P$ (see Definition~\ref{def-TM}).
\label{pan-halt}
\end{definition}
This is much like Cohen's definition of a virus since it characterizes Panopticons
according to their {\em displayed} or {\em manifested} behaviour.
We stress the word {\em execution} in order to preclude situations
where a false alarm is raised for ``Panopticons'' which merely list
actions that are characteristic of Panopticon behaviour without {\em ever} actually invoking them
during their operation. Instead, they operate normally without any
actions taking place that manifest Panopticon behaviour.

Beyond displayed behaviour, however, Panopticons can be reasonably assumed to also
possess {\em deductive} powers, not directly visible or measurable. In other words, one type of such Panopticons may operate by gathering or computing totally new information, distinct from the information already known to it. 
Thus, another type of such Panopticons can
be based on given {\em easily} acquired, or even {\em stolen},
freely provided (in some sense) information. In other words,
based on information the Panopticon acquires for free, in a sense, it deduces further information, perhaps expending some computational effort this time. We model the characteristic Panopticon action, i.e. {\em observation} or {\em surveillance}, using oracle Turing machines, where the freely acquired information is modeled by the oracle set of the machine.
Based on this information, the Turing machine deduces, through its normal computation steps, {\em further} information about its targets.
Below, we describe both these two types of Panopticons.

\begin{definition} (Deductive Panopticons)
A Panopticon is a computer program that by itself or based on observed or stolen (and, thus, acquired without expending computational effort to deduce or produce it) information, deduces (perhaps with computational effort) further information about entities.
\label{pan-oracle}
\end{definition}
In the definition above, the Panopticon operating by itself, i.e. without oracles, is weaker than the one with oracles since the latter is allowed to obtain free advice or information, in the form of an oracle.
Naturally, many other definitions would be reasonable or realistic. Our main motivation behind the ones stated above was a balance of theoretical simplicity and plausibility in order to spark interest on the study on formal properties of Panopticons as well as the difficulty of detecting them algorithmically.

Based on the two formal Panopticon definitions we gave above, we can define the corresponding Panopticon detection problems.
The aim of a Panopticon detection algorithm or Turing machine, is to take as input the encoding of another Turing machine and decide whether it is Panopticon or not based on the formal definition.

\medskip

\noindent{\bf{The Panopticon Detection Problem 1}}

\noindent {\bf Input:} A description of a Turing machine (program).

\noindent {\bf Output:} If the input Turing machine behaves like a Panopticon according to Definition~\ref{pan-halt} output True. Otherwise, output  False.

More formally, if by $L_b$ we denote the language consisting of Turing machine encodings $<M>$ which are Panopticons according to Definition~\ref{pan-halt}, then we want to decide $L_{b}$, i.e. to design a Turing machine that, given $<M>$, decides whether $<M>$ belongs in $L_b$ or not.

\medskip

\noindent{\bf{The Panopticon Detection Problem 2}}

\noindent {\bf Input:} A description of a Turing machine (program).

\noindent {\bf Output:} If the input Turing machine behaves like a Panopticon according to Definition~\ref{pan-oracle} output True. Otherwise, output  False.

\medskip

More formally, if by $L_{d}$ we denote the language consisting of Turing machine encodings $<M>$ which are Panopticons according to Definition~\ref{pan-oracle}, then we want to decide $L_{d}$, i.e. to design a Turing machine that, given $<M>$, decides whether $<M>$ belongs in $L_d$ or not.

\subsection{Behavioral Panopticons}
\label{beh}

Let $Q_{\mbox{\small{pan}}}$ be the set of actions which, when {\em executed}, manifest Panopticon behaviour
(see Definition~\ref{pan-halt}). We will show below that $L_u$ is recursive in $L_b$. This implies
that if we had a decision procedure for $L_b$ then this procedure could also be used for deciding $L_u$ which is undecidable. Thus, no decision procedure exists for $L_b$ too.

\begin{theorem} (Impossibility of detecting behavioural Panopticons) The language $L_b$ is undecidable.
\label{beh}
\end{theorem}

\noindent \textbf{\textit{Proof.}} Our proof is similar to Cohen's proof about the impossibility of detecting viruses.
Let $<M,w>$ be an instance of the Halting problem. We will show how we can decide whether $<M,w>$ belongs in $L_u$ or not using a hypothetical decision procedure (Turing machine) for the language $L_b$. In other words, we will
show that $L_u$ is recursive in $L_b$.

Given $<M,w>$ we design a Turing machine $M^{u-b}$ that modifies the
transition function (see Definition~\ref{def-TM}) of $M$ so as when a final state is reached
(i.e. a state in the set $F$ of $M$) a transition takes place that essentially
starts the execution of the actions in $Q_{\mbox{\small{pan}}}$.
In a sense, $M$ is now a new Turing machine $M'$ containing the actions
of $M$ followed by actions (any of them) described by the states in $P$. Now, $M'$ is given as input the input of $M$, i.e. $w$,
and operates as described above.

Let us assume that there exists a Turing machine $M_b$ that decides $L_b$. Then we can give to it as input $M'$. Suppose that $M_b$ answers that $M' \in L_b$. Since a state in $Q_{\mbox{\small{pan}}}$
was finally activated, as $M_b$ decided, this implies that $M$ halted on $w$ since $M'$ initially simulated $M$ on $w$. Then we are certain that $M$ halts on $w$.

Assume, now, that $M_b$ decides that $M'$ is not a Panopticon. Then 
a state in $Q_{\mbox{\small{pan}}}$ was never invoked,
which implies that no halting state is reached by $M$ on $w$ since a state $Q_{\mbox{\small{pan}}}$
is invoked, in $M'$, only from halting states of $M$, which is simulated by $M'$.
Thus, $M$ does not halt on $w$.

It appears that $M'$ is a Panopticon if and only if $M$ halts on $w$ and, thus, we have shown that $L_u$ is recursive in $L_b$. There is a catch, however, that invalidates this reasoning: if $M$ {\em itself} can exhibit the Panopticon behaviour,
i.e. it can reach a state in $Q_{\mbox{\small{pan}}}$ before reaching a final state.
Then Panopticon behaviour can be manifested without ever $M$ reaching
a final state that would lead $M'$ to invoke a Panopticon state in $Q_{\mbox{\small{pan}}}$, by its construction.
A solution to this issue is to {\em remove}
the states in $Q_{\mbox{\small{pan}}}$ from the transition function of $M$,
giving this new version to $M^{u-b}$ to produce $M'$.
This action would validate the equivalence $M'$ is a Panopticon if and only if $M$ halts on $w$, completing the proof.

More formally, we create a new set of dummy (``harmless'' or ``no-operation'')
``Panopticon'' states $Q^\prime_{\mbox{\small{pan}}}$
which contains a new state for each of the states in $Q_{\mbox{\small{pan}}}$.
%
%
Then we replace the states from $Q_{\mbox{\small{pan}}}$ that appear in the transition function of $M$
with the corresponding states in $Q^\prime_{\mbox{\small{pan}}}$.
Actually, this transformation removes from a potential Panopticon the actions that {\em if} executed would manifest a Panopticon. We stress, again, the fact the mere existence of Panopticon actions
is not considered Panopticon {\em behaviour}.

With this last transformation, $M'$ is a Panopticon if and only if $M$ halts on $w$ and, thus, $L_u$ is recursive in $L_b$.  $\hfill \Box$

\subsection{Deductive Panopticons}
\label{ded}

We, first, prove the undecidability of $S_1$, i.e.
the impossibility of deciding for a given Turing machine (its encoding, to be precise) whether it accepts a language disjoint from a given, fixed, finite set of languages. In other words, it is impossible to detect Turing machines that decide, perhaps with effort, new information sets given some known ones.

\begin{theorem}
The Halting Problem for Turing machines without oracles, i.e. $L_u$, is equivalent to $S_1$.
\label{1st}
\end{theorem}

\noindent \textbf{\textit{Proof.}} We first prove that given an oracle for the $S_1$ we can solve the Halting problem (or, equivalently, recognize the language $L_u$). We construct a Turing machine $M^{S_1}$ such that given $\left\langle M,w\right\rangle $ constructs a Turing machine $M^{'}$ which operates as follows. It ignores its input and simulates, internally, $M$ on $w$. If $M$ accepts $w$, $M^{'}$ accepts its input. Then, $L(M^{'})=\emptyset $ if $M$ does not accept $w$ while $L(M^{'})={\mathit{\Sigma}}^*$ if $M$ accepts $w$. Then, $M^{S_1}$ asks the oracle whether $M^{'}\in S_1$. If the answer is yes, i.e. let $L(M^{'})=\emptyset $, then $M$ does not accept $w$. If the answer is no, then $L(M^{'})={\mathit{\Sigma}}^*$ and, thus, $M$ accepts $w$. We, thus, can recognize $L_u$.

For the other direction, we show that we can recognize $S_1$ given an oracle for the Halting problem (more precisely, $L_u$). We will construct a Turing machine $M^{{'}{'}}$ such that, given $M$, it constructs another Turing machine $M^{'}$ that operates as follows. $M^{'}$ ignores its own input and uses a generator of triples $(i,j,l)$, $1\le l\le k+1$, for simulating the $l$th Turing machine, $M_l$, with $M_{k+1}=M$, on the $i$th string for $l$ steps. Each time one of the Turing machines  $M_1,\ M_2,\dots ,M_k$ accepts a particular input, this fact is recorded on $M^{{'}{'}}${'}s tape. Each time $M_{k+1}$ accepts an input, $M^{{'}{'}}$checks whether the same input was accepted earlier by one of the $M_1,\ M_2,\dots ,M_k$. If no, the process continues. If yes,  $M^{'}$ stops the simulation and $M^{'}\ $accepts its own input. Thus, $L(M){\in S}_1$  if $L(M^{'})=\emptyset $ while $L(M)\mathrm{\notin }S_1$  if $L(M^{'})={\mathit{\Sigma}}^*$, i.e. $M^{'}$ accepts all its inputs, $\varepsilon $ in particular. Then, $M^{{'}{'}L_u}$ may query its oracle set ${\mathrm{L}}_u$ for $\left\langle M^{'},\varepsilon \right\rangle $. If the answer is yes then $M^{{'}{'}}$ rejects $M$, otherwise it accepts it. $\hfill \Box$

\begin{theorem}
The problem of deciding whether $L(M)={\mathit{\Sigma}}^*$\textit{ is equivalent to }$S_2$.
\label{2nd}
\end{theorem}

\noindent \textbf{\textit{Proof.}} We first show that deciding whether\textit{ }$L(M)={\mathit{\Sigma}}^*$\textit{ is recursive in}${\ S}_2$\textit{.} We construct a Turing machine $M^{{{'}{'}{'}S}_2}$ that
takes as input a Turing machine $M$ and constructs from it a Turing machine ${\hat{M}}^{S_1}$, that is a Turing machine with oracle set $S_1$, that operates in the following way. It enumerates strings $x$ over the alphabet $\mathit{\Sigma}$, and for each such string it uses oracle $S_1$ in order  to decide whether $M$ accepts $x$. This can be accomplished in the way described in the first part of the proof of Theorem~\ref{1st}.

Then ${\hat{M}}^{S_1}$ accepts its own input if and only if
a string $x$ is found \textit{not} accepted by $M$, or
\[L({\hat{M}}^{S_1})=\left\{ \begin{array}{c}
\emptyset ,\mathrm{\ if\ }L(M)={\mathit{\Sigma}}^* \\ 
{\mathit{\Sigma}}^*\ \mathrm{otherwise.} \end{array}
\right.\] 
Now $M^{{{'}{'}{'}S}_2}$ asks its oracle $S_2$ whether $L({\hat{M}}^{S_1})\in S_2$, i.e. whether
$L({\hat{M}}^{S_1})$ is disjoint from all sets in $N^{'}$. If the answer is yes, then $L({\hat{M}}^{S_1})=\emptyset $. Consequently, $L(M)={\mathit{\Sigma}}^*$. If the answer is no, on the other hand, then $L({\hat{M}}^{S_1})={\mathit{\Sigma}}^*$ and, thus, $L(M)\neq {\mathit{\Sigma}}^*$. Thus, deciding whether\textit{ }$L(M)={\mathit{\Sigma}}^*$ is recursive in${\ S}_2$\textit{.}

We not turn to showing that ${\ S}_2$ is recursive in the problem of whether $L(M)={\mathit{\Sigma}}^*$. In other words, if by ${\ L}_*$ we denote the codes of the Turing machines which accept all their inputs, then there exists a Turing machine $M^{{'}{'}{'}{'}L_*}$, i.e. a Turing machine with oracle set ${\ L}_*$, which accepts${\ S}_2$.

Given a Turing machine $M^{S_1}$, we define the notion of a \textit{valid computation} of $M^{S_1}$ using oracle $S_1$
in a way similar to notion defined in~\cite{HopUll68,HopUll79}.
A valid computation is a sequence of computation steps such that the next one follows from the current one after a computational (not oracle query) step, according to the internal operation details (i.e. program) of the Turing machine. If a query step is taken, however, i.e. the Turing machine $M^{S_1}$ enters state $q_?$, and the next state is  $q_n$ this means that $M^{S_1}$ submitted a query to the oracle $S_1$ with respect to whether some given Turing machine, say $T$, belongs to the set $S_1$, receiving the answer no. In other words, the oracle replied that $L(T)\mathrm{\notin }S_1$ or, equivalently, $L(T)$ is not disjoint from \textit{all} sets in $N_1$. As evidence for the correctness of this reply from the oracle, we insert a valid computation of the ordinary (i.e. with no oracle) Turing machine $T$ that shows that a {\em particular} string is accepted by, both, $T$ and one of the Turing machines accepting
a language in $N_1$. If, however, after $q_?$ the state $q_y$ follows, no computation is inserted. Intuitively, such a computation would be infinite.

We, now, describe the operation of $M^{{'}{'}{'}{'}L_*}$ with $M^{S_1}$ as input.
Given $M^{S_1}$, $M^{{'}{'}{'}{'}L_*}$ constructs a Turing machine $M^{'}$ to accept \textit{valid computations} of $M^{S_1}$ leading to acceptance, whenever the accepted string, also, belongs to \textit{at least one} of the sets $L^{2}_1,L^{2}_2,\dots , L^{2}_{k}$. The easy case is when the given computation is malformed, when one step does not follow from the previous one according to the internals of the Turing machine, or when the added computation inserted in the $q_?$-$q_n$ case is not valid. In all these cases $M^{'}$rejects its input.

However, there is some difficulty in the $q_?$-$q_y$ case since, as we stated above, there is no obvious \textit{finite} computation evidence for the correctness of the reply. Now the Turing machine $M^{'}$ must decide on its own whether the reply is correct. The reply $q_y$ means that the language accepted by the queried Turing machine $T$ belongs to $S_1$ or, in other words, it is disjoint from all the sets in $S_1$. As in the proof of Theorem~\ref{1st}, $M^{'}$generates all triples $(i,j,l)$, $1\le l\le k+1$, for simulating the $l$th Turing machine, $M^1_1, M^2_2,\dots , M^1_k, M_{k+1}=T$, on the $i$th string for $l$ steps. Each time one of the Turing machines  $M_1,\ M_2,\dots ,M_k$ accepts a particular input, this fact is recorded on  $M^{'}${'}s tape. Each time $T$ accepts an input, $M^{'}$checks whether the same input was accepted earlier by one of the Turing machines $M^1_1, M^1_2,\dots , M^1_k$.
Also, each time one of these machines
accepts an input, $M^{'}$checks whether the same input was accepted earlier by $T$. If none of these two cases apply, the process continues. If one of these two cases, however, holds $M^{'}$ stops the simulation and rejects the computation since it was invalid. It was invalid because a common element was found between the language accepted by $T$ and the language of one the Turing machines accepting languages in $N_1$.

If, however, the computation \textit{is} valid {\em and} it ends at an accepting state for a particular string $x$ which was given as input to $M^{S_1}$, then $M^{'}$ starts generating pairs $(j,l)$, $1\le l\le k$, simulating the $l$th Turing machine, $M^{2}_l$, on $x$ for $j$ steps. If $x$ is accepted by any of these Turing machines, then $M^{'}$ accepts its \textit{own} input.

Based on the above, $M^{'}$ accepts all input strings, that is $L(M^{'})={\mathit{\Sigma}}^*$, if $L(M^{S_1})\mathrm{\notin }S_2$, i.e. when $M^{S_1}$ has valid computations of strings that, also, belong to at least one of the sets $L^{2}_1, L^{2}_2, \dots , L^{2}_{k}$. Otherwise, it accepts the empty set, i.e.  $L (M^{'})=\emptyset $.
Thus, $L(M^{'})={\mathit{\Sigma}}^*$ if and only if $L(M^{S_1}){\mathrm{\notin }S}_2$.

Finally, $M^{{'}{'}{'}{'}L_*}$ asks its oracle whether $L (M^{'})={\mathit{\Sigma}}^*$
or not deciding, in this way, $S_2$ and, thus, detecting deductive Panopticons.   $\hfill \Box$

\section{Weaknesses of formal systems in characterizing Panopticons}
\label{G}

Based on the Recursion Theorem, the following, central to our approach in this Section theorem
is proved in~\cite{HopUll79}:
\begin{theorem}
Given a formal system ${\cal F}$, we can construct a Turing Machine for which
no proof exists in ${\cal F}$ that it either halts or does not halt
on a particular input.
This Turing Machine, denoted by $M_G$, is the following:
\begin{equation} 
g(i,j)= \left\{
\begin{array}{ll}
      1, \mbox{ if there is a proof in \cal F that} f_i(j) \mbox{ is not defined } \\
      \hspace{0.5cm} \mbox{(i.e. does not halt) or,} \mbox{ in other words if there is }\\
      \hspace{0.5cm} \mbox{a proof that the }
        i\mbox{th} \mbox{ Turing Machine does not} \\
      \hspace{0.5cm} \mbox{halt, given input } j \\
      \\
      \mbox{undefined, otherwise}\\
      \end{array} 
\right. 
\label{g}
\end{equation}

\label{impos}
\end{theorem}

We now prove the following, based on Theorem~\ref{impos} and the, effectively
constructible,Turing Machine $M_G$ given in~(\ref{g}):

\begin{theorem} [Imposibility of proving Panopticon status within formal systems]
Let ${\cal F}$ be a consistent formal system.
Then we can construct a Turing Machine for which
there is no proof in ${\cal F}$ that it behaves as a Panopticon
and no proof that it does not behave as a Panopticon, based on Definition~\ref{pan-oracle}
and the set intersection property described in Sections~\ref{contrib} and~\ref{ded}.
\label{imp-formal-proof}
\end{theorem}

\noindent \textbf{\textit{Proof.}}
For some fixed $k$, we define a set $N=\left\{L_1 ,L_2, \dots , L_k \right\}$ of
recursively enumerable languages and a set of corresponding Turing Machines $M_1, M_2, \dots, M_k$
which accept them.
We also assume there exists a recursively enumerable language $L$, accepted by a Turing Machine $M$,
which is disjoint from $N$.

These elements can be effectively constructed. For instance, for $k = 2$,
we can have as $L_1 = \{\mbox{The set of multiples of 2}\},
L_2 = \{\mbox{The set of multiples of 3}\}$, and $L = \{\mbox{The set of primes}\}$. For each
of these languages (in particular $L$) we may construct a corresponding Turing Machine that accepts it.
Given these elements, we proceed as follows.

We construct a Turing Machine $M_?$ which, given $w$ as input, simulates $M$
on $w$ and $M_G$, as defined in~(\ref{g}), on some {\em fixed} input $w_G$, independently of what $w$ is,
by alternating between them in a way similar to the alternation technique applied in Theorem~\ref{1st}.
Then $M_?$ accepts if either of them accepts at some step of the simulation process.

We, now, observe that $L(M_?) = L(M)$,
if $M_G$ does not halt on $w_G$
and $L(M_?) = \Sigma^*$
if $M_G$ halts on $w_G$.
Then, according to the definition
of a deductive Panopticon (see Definition~\ref{pan-oracle} and Section~\ref{contrib}),
$M_?$ is a Panopticon if and only if $M_G$ does not halt on the particular input $w_G$.
But, now, if a proof existed in the formal system ${\cal F}$
that $M_?$ either is a Panopticon or it is not a Panopticon then the same proof could
be used to prove that $M_G$ either halts or does not halt, correspondingly,
contradicting, thus, Theorem~\ref{impos}. $\hfill \Box$

\section{Discussion and directions for future research}

Theorems~\ref{beh},~\ref{1st}, and~\ref{2nd} in Section~\ref{theorems} show
that, even for Panopticons with the simple behaviours described in Definition~\ref{pan-halt} and Definition~\ref{pan-oracle},
it is impossible, in principle, to detect them.
Potential Panopticons, naturally, can have any imaginable, complex, behaviour but then the problem of detecting them may become harder compared to our definitions.

Comparing, now, Theorems~\ref{beh},~\ref{1st}, and~\ref{2nd}, Theorem~\ref{beh} examines the detection of Panopticons based on the execution of {\em specific} visible or detectable actions, i.e. on a {\em behavioural level}, such as connecting to a server and sending eavesdropped information or sending an email to the unlawful recipient. 
Theorems~\ref{1st} and~\ref{2nd} examine Panopticon detection not based on their visible behaviour but from what languages they may accept, without having any visible clue of behaviour or actions, only their {\em descriptions} as Turing machines (i.e. programs or systems). These theorems, that is, examine the detection of Panopticons at a {\em metabehavioural level}.

With respect to the difference between Theorems~\ref{1st} and~\ref{2nd},
we first observe that $L_u$ is recursively enumerable but not recursive while the
$\{<M>|L(M) = \Sigma^*\}$ language is {\em not} recursively enumerable
(see, e.g.,~\cite{HopUll79}). Although they are, both, not recursive (i.e. not decidable),
their ``undecidabilities'' are of different levels, with the $\{<M>|L(M) = \Sigma^*\}$
language considered ``more difficult'' than $L_u$ in restricted types of Turing machines (Panopticons).
For example, the $L_u$ language is decidable for
Context-free Grammars (i.e. for Turing machines modeling
Context-free Grammars) while the $\{<M>|L(M) = \Sigma^*\}$
language is still undecidable. Also, for regular expressions,
the problem of deciding $L_u$ is solvable efficiently (i.e. by polynomial time
algorithms) while the $\{<M>|L(M) = \Sigma^*\}$ language has been shown,
almost certainly, to require exponential time (in the length of the given regular
expression) to solve (see, e.g.,~\cite{HopUll79}).
Therefore, a similar decidability complexity
status is expected from $S_1$ (deductive Panopticons without external advice)
and $S_2$ (deductive Panopticons with external advice in the form of an oracle)
since they are equivalent to the languages $L_u$ and $\{<M>|L(M) = \Sigma^*\}$
respectively. That is, when we consider more restricted definitions of Panopticons
that render the detection problem decidable, then deciding which Panopticons
belong in $S_1$ is expected to be easier than deciding which Panopticons belong in $S_2$.

Finally, Theorem~\ref{imp-formal-proof} shows that
for any formal system ${\cal F}$, we can, effectively, exhibit
a particular Turing Machine for which there is no proof in ${\cal F}$,
that it is either a Panopticon or it is not a Panopticon, emphasizing
the difficulty of recognizing Panopticons by formal means.



As a next step, it is possible to investigate the status of the
Panopticon detection problem under other definitions,
either targeting the behaviour (i.e. specific actions)
of the Panopticon or its information deducing capabilities (e.g. accepting languages
with specific closure properties or properties describable in some formal system such as second order logic).
Our team plans to pursue further Panopticon definitions in order to
investigate their detection status, especially for the
decidable (and, thus, more practical) cases of suitably constrained Panipticons.

In conclusion, we feel that the formal study of the power and limitations
of massive surveillance establishments and mechanisms
of today's as well as of the future Information Society can be, significantly, benefitted
from fundamental concepts and deep results of computability and computational complexity theory.
We hope that our work will be one step towards this direction.

\end{document}